\documentclass[fleqn,usenatbib]{mnras}
\usepackage[normalem]{ulem}

\usepackage{newtxtext,newtxmath}

\usepackage[T1]{fontenc}
\usepackage{ae,aecompl}

\usepackage{graphicx,xspace}

\newcommand{\eqn}[2][]{Equation#1~\ref{eq:#2}} 
\newcommand{\fig}[2][]{Figure#1~\ref{fig:#2}}

\newcommand{\sect}[2][]{Section#1~\ref{sec:#2}}
\newcommand{\app}[2][]{Appendix#1~\ref{sec:#2}}
\newcommand{\U}[1]{\ensuremath{\mathrm{~#1}}}     
\newcommand{\yr}{\U{yr}}
\newcommand{\Myr}{\U{Myr}}          
\newcommand{\Gyr}{\U{Gyr}}          
\newcommand{\pc}{\U{pc}}
\newcommand{\kpc}{\U{kpc}}
\newcommand{\Mpc}{\U{Mpc}}          
\newcommand{\Msun}{\U{M}_{\odot}\xspace}

\newcommand{\cc}{\U{cm^{-3}}}

\newcommand{\kms}{\U{km\ s^{-1}}}
\newcommand{\erg}{\U{erg}}

\newcommand{\kick}{{\tt runaways}\xspace}
\newcommand{\nokick}{{\tt no\,runaways}\xspace}

\newcommand{\ramses}{{\small RAMSES}\xspace}
\newcommand{\starburst}{{\small STARBURST99}\xspace}
\newcommand{\sunrise}{{\small SUNRISE}\xspace}
\newcommand{\agora}{{\small AGORA}\xspace}

\newcommand{\lund}{Department of Astronomy and Theoretical Physics, Lund Observatory, Box 43, SE-221 00 Lund, Sweden}

\title[Runaways mimic star formation]{Runaway stars masquerading as star formation in galactic outskirts}

\author[E. P. Andersson et al.]{
Eric P. Andersson\thanks{E-mail: eric@astro.lu.se},
Florent Renaud
and Oscar Agertz
\\
\lund \\
}

\date{Accepted XXX. Received YYY; in original form ZZZ}

\pubyear{2020}

\begin{document}
\label{firstpage}
\pagerange{\pageref{firstpage}--\pageref{lastpage}}
\maketitle

\begin{abstract}
In the outskirts of nearby spiral galaxies, star formation is observed in extremely low gas surface densities. Star formation in these regions, where the interstellar medium is dominated by diffuse atomic hydrogen, is difficult to explain with classic star formation theories. In this work, we introduce runaway stars as an explanation to this observation. Runaway stars, produced by collisional dynamics in young stellar clusters, can travel kilo-parsecs during their main sequence life time. Using galactic-scale hydrodynamic simulations including a treatment of individual stars, we demonstrate that this mechanism enables the ejection of young massive stars into environments where the gas is not dense enough to trigger star formation. This results in the appearance of star formation in regions where it ought to be impossible. We conclude that runaway stars are a contributing, if not dominant, factor to the observations of star formation in the outskirts of spiral galaxies.
\end{abstract}





\section{Introduction}
\label{sec:intro}
The relationship between star formation rate (SFR) density $\Sigma_{\rm SFR}$ and gas surface density $\Sigma_{\rm g}$, commonly referred to as the star formation (SF) relation, was suggested to follow a power-law by \citet{Schmidt1959}. The canonical SF relation is typically quoted with a slope of 1.4 with a break appearing at a critical threshold \citep[see e.g.][]{Kennicutt1989,Kennicutt1998,Kennicutt&Evans2012}. The break occurs at $\sim\!10\Msun\,{\pc}^{-2}$ and is attributed to the transition between molecular hydrogen H$_2$ and neutral atomic hydrogen HI \citep{Wong&Blitz2002,Kennicutt+2007,Bigiel+2008,Bolatto+2011}. However, the underlying reasons of the transition are debated (see e.g. \citealt{Schaye2004,Krumholz+2005,Krumholz+2009,Renaud+2012,Federrath2013} or \citealt{Krumholz2014} for a review). 

In the outskirts of spiral galaxies and dwarf irregular galaxies, the SF relation extends into extremely diffuse gas going from $\Sigma_{\rm g}\sim 10\Msun\,\yr^{-1}$ toward $\Sigma_{\rm g}\sim 1\Msun\,\yr^{-1}$ \citep{Roychowdhury+2009,Bigiel+2010, Bolatto+2011,Elmegreen&Hunter2015} in which SF proceeds extremely slowly with a roughly constant depletion time of $100\Gyr$. \citet{Elmegreen2015,Elmegreen2018} found that a SF relation with a slope of 2 for the outer galaxy if the disc flares, i.e. if the thickness is regulated by gas self-gravity and a radially uniform velocity dispersion. \citet{Krumholz2013} suggested another model, in which star formation can occur in an atomic medium with a separate cold and warm phase. \citeauthor{Krumholz2013} argued that in regions with low star formation rate (e.g. galactic outskirts), the transition between HI and cold star forming H$_2$ is mediated by hydrostatic balance. SF then proceeds slowly, with depletion times of $\sim100\Gyr$, in agreement with observations \citep[e.g.][]{Bigiel+2010,Bolatto+2011}. Here we show that runaway stars, formed in dense gas and ejected into low density regions, naturally explain the observed third regime of SF in galactic outskirts.

Runaway stars are produced by close encounters and binary disruption due to stellar evolution in young stellar clusters \citep{Blaauw1961,Poveda+1967}. These stars have been studied extensively, both observationally \citep[e.g.][]{Gies&Bolton1986,Gies1987,Stone1991,Hoogerwerf+2000,Silva&Napiwotzki2011,Appelaniz+2018,DorigoJones2020,Raddi+2020} and through modeling \citep[e.g.][]{Ceverino&Klypin2009,Eldridge+2011,Moyano+2013,Oh&Kroupa2016,Kim&Ostriker2018,Andersson+2020}. Typically $5-10$ per cent of massive OB-type stars have velocities exceeding $30\kms$ and can travel hundreds of pc to several kpc before exploding as core-collapse supernovae (SNe). Moreover, the less massive B stars ($\sim\!4\Msun$) are more numerous and can travel significantly further due to their longer lifetimes ($\sim150\Myr$). As such, they can reach the galactic outskirts and contribute to the observational tracers of the SF activity, yet without direct physical connection to the formation sites. By expanding on the results of \citet{Andersson+2020}, we show in this Letter that this mechanism yields an observable signature in striking agreement with SF in galactic outskirts, where gas surface densities are extremely low, as observed by \citet{Bigiel+2010}.

\section{Numerical setup}\label{sec:isodisc_sim_setup}
This work uses the two isolated Milky Way-like galaxies described in \cite{Andersson+2020}. We compare one (referred to as \kick), which includes runaway stars where individual stars are tracked both in terms of stellar evolution and kinematically, to an identical simulation ignoring runaway stars (referred to as \nokick). We briefly describe the numerical method here, and refer to \cite{Andersson+2020} for details.

We ran the two simulations for $250\Myr$ using the $N$-body + Adaptive Mesh Refinement (AMR) code \ramses \citep{Teyssier2002}, which treats dark matter and stars as collisionless particles and computes the fluid dynamics on a grid with adaptive resolution assuming ideal mono-atomic gas with adiabatic index $\gamma=5/3$. Gas cooling is metallicity-dependent and treated using tabulated values. SF is controlled by a density threshold ($100\cc$) with the SFR density computed from the cell gas density divided by the local free-fall time and scaled with an efficiency of $5$ per cent. The details of this method are discussed in \citet{Agertz+2013}. The resolution of the grid follows a quasi-Lagrangian refinement strategy for which a cell is refined if it contains more than 8 particles, or more than $4014\Msun$ of baryonic matter, down to a spatial resolution of $9\pc$. The initial conditions are the same as those used for the isolated disc in the \agora project \citep{Kim+2014,Kim+2016} and gives a galaxy similar to the Milky Way, but with a gas fraction of $20$ per cent.

Star particles are initially sampled with a mass resolution of $500\Msun$ and immediately split into two groups using the initial mass function (IMF) from \citet{Kroupa2001}: i) low mass stars (LMS; $<8\Msun$) are grouped and represented by a star particle for which we consider mass loss, Fe and O enrichment as well as momentum and energy injection from type Ia supernovae (SNe) and asymptotic giant branch winds \citep[see][for details]{Agertz+2013}; ii) high mass stars (HMS; $\geq8\Msun$) are treated as individual stars with a feedback model accounting for fast winds and core-collapse SNe. The mass-loss rate from the fast winds is computed with a modified version of the model by \citet{Dale&Bonnell2008} and depends on stellar mass and metallicity. Core-collapse SNe occurs when HMS leave the main sequence and results in the injection of $10^{51}\erg$ of energy in the gas. The main sequence time is computed with the age-mass-metallicity fit by \cite{Raiteri+1996}. In cases when the Sedov-Taylor phase of the SN is unresolved (resulting in problems with a self-consistent development of the momentum build up during this phase), we explicitly inject the associated momentum using the method from \cite{KimOstriker2015}. This ensures that we capture the effect of SNe even in region with lower resolution. 

In the \kick simulation, we add natal velocity kicks to the HMS sampled from the power-law distribution derived from $N$-body simulations of clusters by \citet{Oh&Kroupa2016}:
\begin{equation}\label{eq:v_dist}
    f_v \propto v^{-1.8},\quad v \in [3,385]\,\kms.
\end{equation}
This results in $\sim 14$ per cent of stars being runaway\footnote{We define runaway stars as stars with peculiar velocities $>30\kms$. For a discussion on this value, see \citet{Andersson+2020} and references therein.} with a mean velocity of $90\kms$.

\subsection{Low mass runaway stars}\label{sec:LMS_resample}
The model in \citet{Andersson+2020} limits the runaway mechanism to massive ($>8\Msun$) stars because these stars are responsible for the majority of stellar feedback. In this work, where we focus on tracing the resolved SF relation discussed in \sect{intro}, we extend the model to sample individual stars in the mass range $4-100\Msun$ in order to account for the non-negligible contribution to far ultraviolet (FUV) emission of low mass B-stars. These stars are an important contributor to our SF tracer (see \sect{SF_FUV}). This extension is achieved by re-sampling the mass\footnote{Assigning new masses inevitably leads to a different stellar evolution. This implies a discrepancy between the stellar evolution of the analysed stars and those evolved in the simulation. This is a limitation of our model, which we take into consideration when drawing conclusions from our results.} of every HMS in the aforementioned mass range using the Kroupa IMF as a post processing step. For technical reasons, we focus our analysis on $\sim150\Myr$ of evolution of the galaxy, which roughly corresponds to the main sequence life time of a $4\Msun$ star. We therefore use this mass as a lower limit, resulting in a mass range between $4-100\Msun$ for the HMS. We keep \eqn{v_dist} (which does not have any mass dependence) as the velocity distribution. The re-sampling increases the stellar mass in the HMS population by $1.4\times10^{7}\Msun$. We remove mass from LMS corresponding to stars in the mass range $4-8\Msun$ since stars in this mass range are now included as HMS, resulting in a decrease in total mass of $1.7\times10^{7}\Msun$. The discrepancy between the two comes from mass loss due to stellar evolution which is unaccounted for by this re-sampling.

\section{Results}\label{sec:results}
\subsection{The resolution dependence of the star-formation relation}

\begin{figure*}
    \centering
    \includegraphics[width = 0.92\columnwidth]{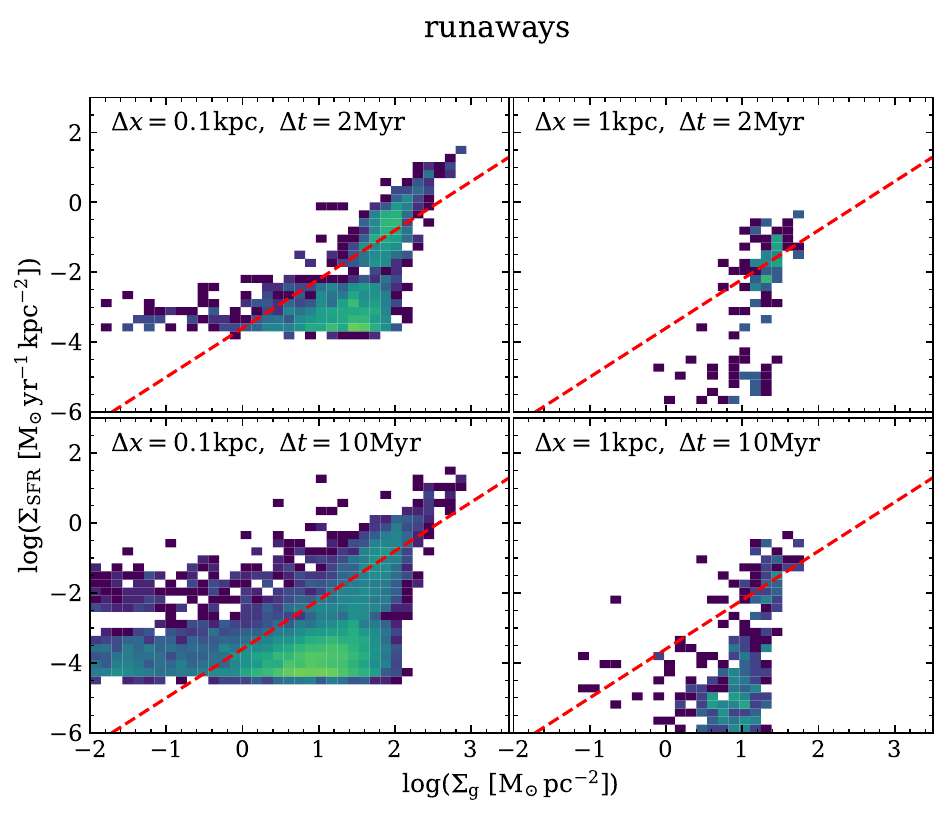}
    \includegraphics[width = 0.92\columnwidth]{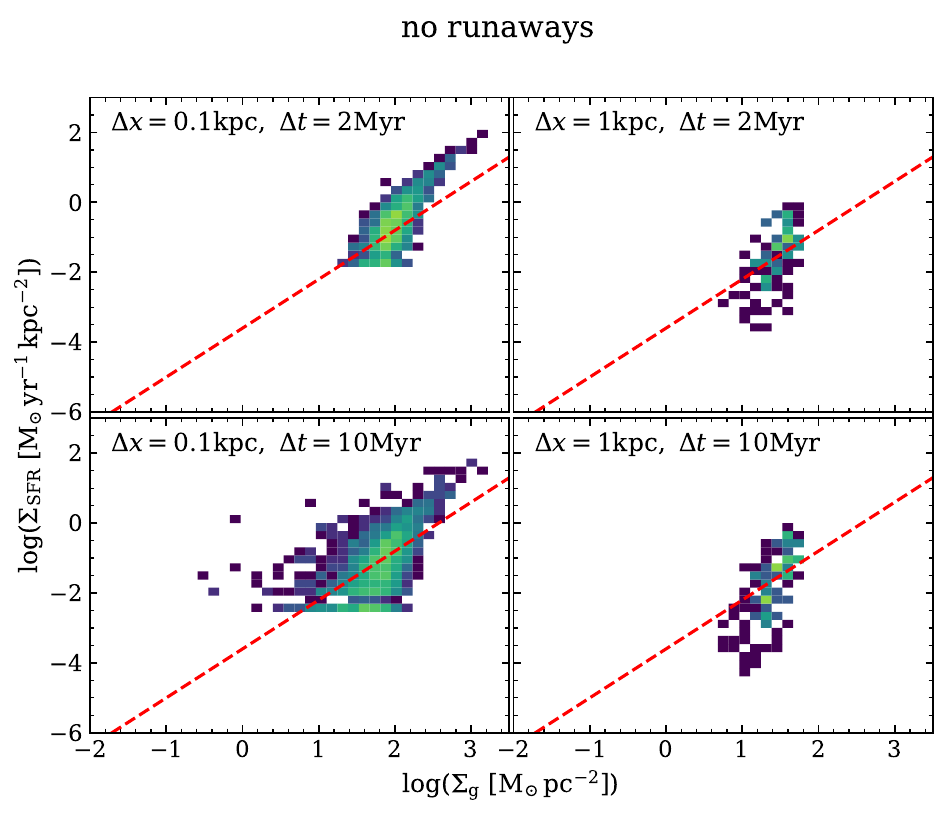}
    \caption{SFR surface density as function of gas surface density for different resolution in space ($\Delta x$) and time ($\Delta t$) for the model with runaway stars (left) and without (right). Small values for $\Delta x$ and $\Delta t$ results in the tightest coupling between stars and gas. To guide the eye, the canonical SF relation with a slope of 1.4 \citep{Kennicutt1989,Kennicutt1998} is shown by the red dashed line.}
    \label{fig:KS_grid}
\end{figure*}

At the end of the simulation, we derive the local SFR surface density by considering the mass of all stars with an age less than $\Delta t$ in square bins with sides $\Delta x$ placed in a uniform grid on the face-on view of the galaxies. We compare this to the gas surface density in the same bins and show the result in \fig{KS_grid} for different choices of $\Delta t$ ($0.1\kpc$ and $1\kpc$) and $\Delta x$ ($2\Myr$ and $10\Myr$). We find that both galaxies follow the empirical relation $\Sigma_{\rm SFR}\propto\Sigma_{\rm g}^{1.4}$ \citep[][red dashed line in \fig{KS_grid}]{Kennicutt1998} at high gas densities, with a break at $\sim10-100\Msun\,{\rm pc}^{-2}$ going into the regime of slow SF\footnote{We use the terms slow/fast SF to describe trends following long/short depletion times (i.e. lines of constant $\tau_{\rm dep}=\Sigma_{\rm g}/\Sigma_{\rm SFR}$). This is sometimes referred to as SF efficiency (then defined as the inverse of depletion time). In this work, we reserve the term efficiency to describe the conversion of gas mass into stellar mass (without concern for the timescale). For a more detailed discussion on differences between SF efficiency and depletion time, see e.g. \citet{Semenov2018,Renaud+2019}}. Our galaxies show slightly faster SF at high surface densities ($\Sigma_{\rm g}\gtrsim100\Msun\pc^{-2}$), albeit being withing the typical observed scatter in local spiral galaxies \citep[see e.g.][]{Wong&Blitz2002,Crosthwaite&Turner2007,Kennicutt+2007,Schuster+2007,Bigiel+2008}. We find that increasing the resolution in time and space, i.e., decreasing $\Delta t$ and $\Delta x$, results in less dispersion in the SF relation for both \kick and \nokick. We attribute this to a tighter correlation between newly formed stars and their natal gas clouds. Increasing $\Delta t$ causes the scatter to increases because of the decoupling between stars and gas due to e.g. stellar feedback, dynamical drift, cloud dissolution. These different decoupling mechanisms have a range of spatial scales and time scales. By increasing $\Delta x$, the variations of the gas density from region to region are averaged out, thus reducing the scatter in \fig{KS_grid}. However, increasing the temporal and spatial scales inevitable causes the measurements to deviate from the relation imposed from the \emph{local} (cell-based) SF law. Similarly to our results, \citet{Khoperskov&Vasiliev2017} found that on small spatial scales ($\lesssim100\pc$) the SF relation as measured from far ultraviolet (FUV) flux deviates from that estimated by free-fall collapse of molecular clouds \citep[see e.g. equation 21 in ][]{Krumholz+2005}. This implies that on such scales the SF relation reflects the various evolutionary stages of individual star forming clouds, hence the relation is lost as clouds are destroyed or stars escape \citep[see also][]{Onodera+2010}.

Naturally, we find more scatter in the SF relation in the \kick model, since the velocity kicks amplify the decoupling of the runaway stars from their natal gas. This masquerades as star formation activity in regions of low $\Sigma_{\rm g}$, as best seen in left panels at $-1\lesssim\log(\Sigma_{\rm g})\lesssim0$ and $\log(\Sigma_{\rm SFR})\sim-2$.

In all panels of \fig{KS_grid}, there is a floor in $\Sigma_{\rm SFR}$ (most clearly visible for $\Delta x=0.1\kpc$). This floor is a result of having a single star particle within a bin of size $\Delta x$, and is therefore set by the finite resolution in stellar masses. In \kick the resolution is $4\Msun$, while in the \nokick, it is the mass of the entire unresolved stellar population. As discussed earlier, increasing $\Delta t$ allows stars to travel further, reaching a larger range of densities. For very long time scales ($\sim100\Myr$), runaway stars reach the outskirts of the galaxy, as discussed in the remainder of \sect{results}. 

\subsection{Runaway stars explain observations of low SFR}\label{sec:SF_FUV}

\begin{figure*}
    \centering
    \includegraphics[width = 1.95\columnwidth]{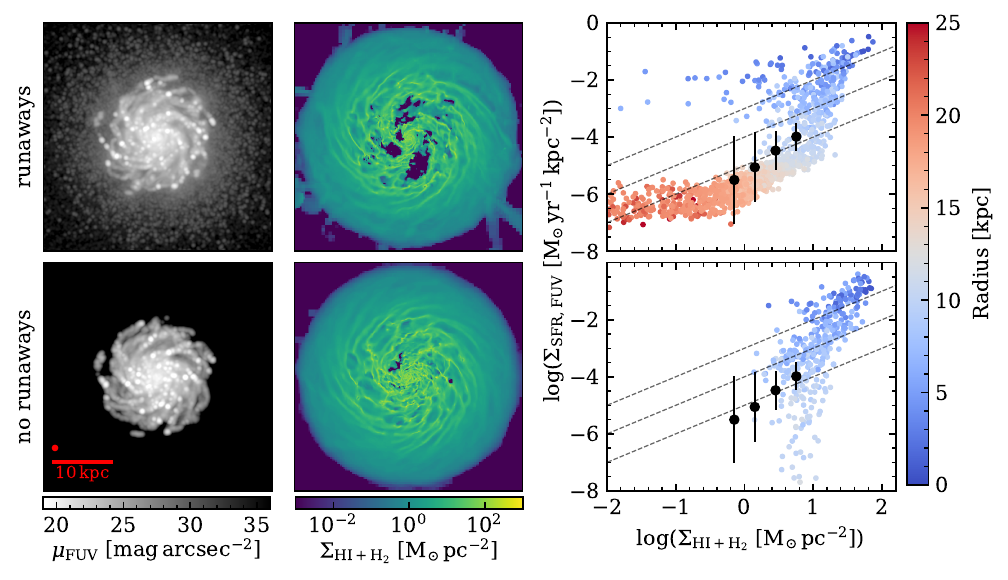}
    \caption{{\it Left:} FUV surface brightness maps derived by assigning spectra to each star (see text for details). Each pixel has been convolved with a Gaussian filter with FWHM of 4" converted to physical units by assuming a distance of $20\Mpc$ to the galaxies. The point spread function is shown by the red circle. {\it Center:} Surface density of HI and H$_2$ gas along the line of sight. The scale is the same as for that in the left panel. {\it Right}: Resolved SFR density as function of gas surface density colour-coded by distance to the centre of the galaxy. $\Sigma_{\rm SFR,FUV}$ is computed from FUV intensity (\eqn{SFR_FUV}), in $1\kpc$ squares. The colour-coding clearly shows that the low-$\Sigma_{\rm SFR}$ feature is a radial trend related to the inclusion of runaway stars. The black dots with errorbars shows the mean and scatter for the resolved KS-relation observed in outer regions of spiral galaxies by \citet{Bigiel+2010}. The dashed lines corresponds to constant depletion times of $100\Gyr$, $10\Gyr$ and $1\Gyr$ from bottom to top respectively.}
    \label{fig:FUV_KS}
\end{figure*}

To avoid arbitrary time scales ($\Delta t$), we create mock observations of the FUV flux. These are shown in \fig{FUV_KS} as surface brightness maps (left panel) from which we derive the SFR density. In the right panels, we plot them against the surface density of neutral and molecular hydrogen $\Sigma_{\rm HI+H_2}$ (centre panels). This accounts for the dimming of FUV luminosity due to stellar evolution, thus introduces a self-consistent timescale, and lifts the requirement for an arbitrary $\Delta t$. This measurement of SFR is therefore consistent with that estimated observationally. \app{generate_SSP} details how we produce and observe the mock spectra. The SFR is then computed as
\begin{equation}\label{eq:SFR_FUV}
  {\rm SFR}\ [{\rm M}_{\odot}\,{\rm yr}^{-1}] = 0.68\times10^{-28}\ I_{\rm FUV}\ [{\rm erg}\,{\rm s}^{-1}\,{\rm Hz}^{-1}],
\end{equation}
where $I_{\rm FUV}$ is the FUV intensity integrated over the GALEX-FUV filter. We calibrate the $I_{\rm FUV}$ such that the global SFR is the same as that measured in the simulation, as detailed in \app{generate_SSP}. \eqn{SFR_FUV} is identical to that derived by \citet{Salim+2007} and later adopted by \citet{Leroy+2008} and \citet{Bigiel+2010}. Note that this is the unobstructed SFR (not accounting for the contribution of embedded SF observed in infrared re-emission), and is therefore a lower limit of the SFR. However, our main finding is the feature in the low gas density regime, where the SFR densities are largely unaffected by extinction.

As shown in the top right panel of \fig{FUV_KS}, we find a radial dependence in the branch at low $\Sigma_{\rm g}$ and low $\Sigma_{\rm SFR}$. In fact, we find that the transition into this feature corresponds to the regions outside of the star forming disc ($\gtrsim10\kpc$), thus explaining the absence of this branch in the \nokick simulation. The signal arises because the runaway mechanism ejects stars into regions where the density is too low for SF to be active. This creates the illusion of SF in gas with extremely low density, seen as the aforementioned feature extending from $\Sigma_{\rm SFR}\sim10^{-4}\Msun\yr^{-1}\kpc^{-2}$, toward the bottom left. Initially, the branch roughly extends along a line of constant depletion time (shown by dashed the lines) corresponding to extremely slow star formation ($\tau_{\rm dep}=100\Gyr$), and flattens out as it reaches very low $\Sigma_{\rm HI+H_2}$ ($\leq0.1\Msun\,\pc^{-2}$). A comparison to measurements of SF in outer regions of observed spiral galaxies \citep{Bigiel+2010} reveals a striking similarity to the feature produced by our runaway model. How far out the branch extends radially depends on the runaway model, i.e. the velocity distribution and the mass distribution of the stars, as discussed in \sect{discussion}.

The middle panels of \fig{FUV_KS} compare the gas structure between \kick and \nokick. The \kick (top) simulation features large under-dense regions within the star forming disc (a few kpc from the centre). The repeated transport of runaway stars into low density medium (e.g. inter-arm) allows these feedback bubbles to expand to large volumes and survive for longer timescales compared to the \nokick case where they are confined to dense media. As in the galactic outskirts ($\log(\Sigma_{\rm SFR})\sim-6$), these low gas surface density regions at $\log(\Sigma_{\rm SFR})\sim-2$ harbour an unexpected SF activity introduced by the presence of runaway stars. Contrary to the galactic outskirts, these measurements are indicative of fast SF ($\tau_{\rm dep}\sim10\Myr$) as seen in the top right of \fig{FUV_KS}, because runaway stars are more abundant in the inner parts of galactic disc.

\section{Discussion}\label{sec:discussion}
The physical process responsible for the observational detection of young stars at gas surface densities as low as $1\Msun\,{\pc}^{-2}$ in resolved galaxies \citep[e.g.][]{Bigiel+2008,Wyder+2009,Elmegreen&Hunter2015,Bigiel+2010} has been debated \citep[][for a review]{Krumholz2014}. \citet{Elmegreen2015} argued that the longer depletion times in the outskirts arises from disc flaring reducing the local volume density of gas, while keeping the surface density relatively high. This is in contradiction with our \nokick model, were such a feature should be visible. \citet{Krumholz2013} suggested that the trend at low $\Sigma_{\rm SFR}$ can be explained by inefficient star formation in molecule-poor gas in the outskirts of galaxies. A key property of their model is the background interstellar radiation, which we do not account for in the star formation law in our simulations. Therefore, we cannot not rule out low levels of star formation in molecule-poor gas in the galactic outskirts. Nevertheless, runaway stars, as modelled in this work, must contribute to the observed SF signal, at least to some extent. Fully understanding the contribution from runaway stars likely requires advanced models with full $N$-body treatment of all stellar clusters, including consistent descriptions of the clusters natal properties, such as binary fraction. Furthermore, this needs to be accounted for over cosmological times to obtain a self-consistent local FUV background. This is beyond the scope of current models. 

Star formation in very diffuse gas could either reflect {\it in-situ} star formation, or result from rapid migration of stars, as we advocate for here. As discussed above, the former case calls for an additional regime of star formation. However, the two possibilities would lead to a different stellar mass function. By comparing FUV flux with H$\alpha$ emission from HII regions, \citet{Meurer+2009} found a deficit of massive O stars associated with the FUV bright outskirts \citep[see also][]{Werk+2010}. Note that there is a debate surrounding their conclusions \citep[see e.g.][]{Fumagalli+2011,Andrews+2013,Andrews+2014}. Unless arguing for a non universal IMF \citep[e.g.][]{Pflamm-Altenburg&Kroupa2008}, a deficit in O stars supports our scenario since we find that the main contributor to the FUV signal are stars in the mass range $5-7\Msun$, with almost no contribution from O stars with mass $>20\Msun$.

In this work, we apply a model which treats stars as individual particles on galactic scales, which is numerically challenging. In order to reduce computational cost, the model in \citet{Andersson+2020} is limited to massive stars ($>8\Msun$) because these trace the majority of stellar feedback. In this work, we have extended the model to include individual stars down to $4\Msun$, thus accounting for their contribution to the FUV flux in galaxies\footnote{Although the contribution to FUV intensity is limited for low mass stars, their long main sequence lifetime makes them important for the radial dependence of the SF relations.}. The velocity distribution (\eqn{v_dist}) is that of all stars in a natal cluster, implying that in principle we can choose to re-sample for any stellar masses. However, because of mass segregation and the dynamics of many-body interactions, the velocity distribution is known to be mass dependent. More massive stars are more likely to receive stronger kicks. \cite{Appelaniz+2018} estimated that in the field $10-12$ per cent of O stars are runaways, while this is fraction is reduced to $\sim6$ per cent for B stars \citep[see also][]{Eldridge+2011}. In future work, we will investigate how our results depend on varying the distribution of kick velocities. 

\section{Summary \& Conclusions}
Using hydrodynamic simulations of an isolated Milky Way-like galaxy \citep{Andersson+2020}, we show how runaway stars change the appearance of SF, as seen in the $\Sigma_{\rm g}$-$\Sigma_{\rm SFR}$ plane. We demonstrate how the SF relation depends sensitively on the spatial and temporal scales over which they are averaged. This sensitivity is increased by runaway stars, since their high velocities implies that they quickly leave their natal environments. By estimating the SFR from the FUV intensity, we remove the necessity of choosing an \emph{ad hoc} timescale and produce a SF relation consistent with that derived from observations.

Our main result is a feature in the SF relation at $\Sigma_{\rm SFR}\sim10^{-4}-10^{-6}\Msun\,\yr^{-1}\,{\kpc}^{-2}$ in low surface density gas, with a galactocentric radial dependence, found exclusively in our model including runaway stars. This feature is in excellent agreement with that observed in the outer regions of spiral galaxies \citep{Bigiel+2010}. We show that this feature arises by ejecting massive FUV emitting stars (via the runaway mechanism) from star formation regions into low-density gas. This results in the presence of young stars in gas with densities that are too low to trigger star formation. Therefore, it produces an unexpected signature of SF, with a strong radial dependence.

In conclusion, we argue that the SF relation in the outer regions of spiral galaxies is produced by a small, albeit observable, population of individual stars formed in denser environments and transported there by the runaway mechanism. Although our model can not rule out star formation in atomic gas \citep{Krumholz2013}, runaway stars is at the very least a contributing, if not dominant, factor to establishing the SF relation in outer regions of galaxies. 

\section*{Acknowledgements}
We thank the anonymous referee for comments which improved this work. We also express gratitude for the stimulating discussions at the 2020 Ringberg Virtual Seminar Series. EA acknowledges discussions with Mark Krumholz. We acknowledge support from the Knut and Alice Wallenberg Foundation, the Swedish Research Council (grant 2014-5791) and the Royal Physiographic Society of Lund. We used computational resources at LUNARC hosted at Lund University, on the Swedish National Infrastructure for Computing (SNIC 2018/3-649), as well as allocation LU 2019/2-27. 

\section*{Data Availability}
The data underlying this article will be shared on reasonable request to the corresponding author.



\bibliographystyle{mnras}
\bibliography{ref}



\appendix
\section{Generating synthetic spectra for stellar populations}\label{sec:generate_SSP}
The mock observations used in this work are derived by combining a wave propagation method similar to that used in \sunrise \citep{Jonsson2006} with stellar spectra from a modified version of the stellar population synthesis code \starburst \citep{Leitherer+1999,Vazquez&Leitherer2005,Leitherer+2010,Leitherer+2014}. All HMS use evolving spectra for individual stars while the LMS uses an evolving spectra from the stellar population with stars in the relevant mass range. The spectra of each source is propagated through the gas at the highest AMR resolution ($\sim9\pc$) to the observer. We account for extinction using a dust attenuation curve from \citet{Li&Draine2001} assuming a uniform dust-to-gas ratio of 0.01. We compute the photometric intensity in the GALEX-FUV band for the stars which formed during the simulation (i.e. younger than $250\Myr$). To account for missing of FUV flux from unresolved populations and loss of mass due to sampling new masses for HMS particles (see \sect{LMS_resample}), we artificially boost $I_{\rm FUV}$ such that the global SFR derived from \eqn{SFR_FUV} matches that from the simulation. Furthermore we apply a Gaussian filter with FWHM of 4" (i.e. standard deviation of $\sigma=1.7$") to all pixels to simulate the angular resolution of GALEX satellite. To convert to physical units we assume a distance of $20\Mpc$. The surface brightness maps were computed from the magnitudes using 
\begin{equation}
    m_{\rm FUV} = -2.5\log\left(\frac{f_{\nu}}{\erg\,{\rm s}^{-1}\,{\rm cm}^{-2}\,{\rm Hz}^{-1}}\right) - 48.6,
\end{equation}
where $f_{\nu}$ is the spectral flux density in the GALEX-FUV band. When we compute $f_{\nu}$ we assume an initial distance of $10\pc$ and add a distance modulus corresponding to $20\Mpc$. For the conversion to surface brightness, we compute the pixel angular size at a distance of $20\Mpc$.


\bsp	
\label{lastpage}
\end{document}